\documentclass[10pt,preprintnumbers,showpacs,amsmath,amssymb,cite,onecolumn]{revtex4}
\input epsf.sty
\usepackage{graphicx}
\usepackage{txfonts}
\usepackage{bm}   
\usepackage{amsmath,mathrsfs}
\usepackage{multirow}
\begin{document}

\title{Valence quark distributions of the proton from maximum entropy approach}
\author{
Rong Wang$^{1,2,3}$\email{rwang@impcas.ac.cn},
Xurong Chen$^{1}$
}
\affiliation{
$^1$ Institute of Modern Physics, Chinese Academy of Sciences, Lanzhou 730000, China\\
$^2$ Lanzhou University, Lanzhou 730000, China\\
$^3$ University of Chinese Academy of Sciences, Beijing 100049, China\\
}
\date{\today}

\begin{abstract}
We present an attempt of maximum entropy principle to determine valence
quark distributions in the proton at very low resolution scale $Q_0^2$.
The initial three valence quark distributions are obtained with limited
dynamical information from quark model and QCD theory. Valence quark
distributions from this method are compared to the lepton deep inelastic
scattering data, and the widely used CT10 and MSTW08 data sets. The obtained
valence quark distributions are consistent with experimental observations
and the latest global fits of PDFs. Maximum entropy method is expected
to be particularly useful in the case where relatively little information
from QCD calculation is given.
\end{abstract}
\pacs{12.38.-t, 14.20.Dh, 13.15.+g, 13.60.Hb}

\maketitle

\section{Introduction}
\label{SecI}

Determination of parton distribution functions (PDFs) of the proton is of
high interest in high energy physics \cite{CTEQ6,CT10,MSTW,GRV98,ABM},
as PDFs are an essential tool for standard model (SM) phenomenology,
theoretical prediction study and new physics search.
In perturbative quantum chromodynamics (QCD) theory, factorization allows
for the computation of the hard partonic scattering processes involving initial
hadrons, which requires the knowledge of the PDFs in the nucleon. The widely
used PDFs are extracted from global QCD analysis of experimental data on
deep inelastic scattering (DIS), Drell-Yan (DY) and jet production processes.
The initial parton distributions at low scale $Q_0^2$ are called the
nonperturbative input. Valence quarks are the main part of the nonperturbative
input, for they carry most of the momentum of the proton. In the global analysis,
the nonperturbative input is parameterized and evolved to high $Q^2$
to fit with the experimental measurements.

So far, the nonperturbative input cannot be calculated in theory,
due to the complexity of nonperturbative QCD. However,
there are many calculations of valence quark distributions from
models, such as MIT bag model \cite{Steffens,RevValence} and
the Nambu-Jona-Lasinio model \cite{Mineo}.
These model-calculated valence quark distributions in the nucleon
are in agreement with global analysis.
Determination of the nonperturbative input not from the global fit
procedure is not only a complementary to current extraction of PDFs,
but also helps us understand the structure and nature of the hadrons.
In addition, precise determination of valence quark distributions is important
for detailed study of sea quarks in intermediate $x$ region \cite{Chang2014}.

In this article, we try to determine the valence quark distributions of the
proton using maximum entropy method, based on some already known structure
information and properties of the proton in the naive quark model and QCD theory.
The maximum entropy principle is a rule for converting certain types of information,
called testable information, to a probability assignment
\cite{Jaynes1,Jaynes2,Caticha,Toussaint}.
In this analysis, the known properties of the proton are the testable information;
and the valence quark distributions are the probability density functions need
to be assigned. Maximum entropy method gives the least biased estimate
possible on the given information. It is widely used in Lattice QCD (LQCD)
\cite{Asakawa,Ding}, with reliable results and high efficiency.

The organization of the paper is as follow. A naive nonperturbative input
is introduced in Section \ref{SecII}. Section \ref{SecIII} discusses the
standard deviations of parton momentum distributions, which are related to
the quark confinement and Heisenberg uncertainty principle.
In Section \ref{SecIV}, the maximum entropy method is demonstrated.
Section \ref{SecV} presents comparisons of our results with experimental
data and the global analysis results. Finally, discussions and summary
are given in Section \ref{SecVI}.

\section{A naive nonperturbative input from quark model}
\label{SecII}

Quark model is very successful in hadron spectroscopy study and describing
the reaction dynamics. Quark model is based on some basic symmetries,
which uncovers some important inner structures of the hadrons.
The proton consists of a complex mixture of quarks and gluons
in hard scattering processes at high $Q^2$.
In the view of quark model, the origin of PDFs are the three valence quarks.
In the dynamical PDFs model, the sea quarks and gluons are radiatively generated
from three dominated valence quarks and ``valence-like" components
which are of small quantities \cite{GRV95,GRV98,PRD79}.

The solutions of the QCD evolution equations for parton distributions
at high $Q^2$ depend on the initial parton distributions
at low $Q^2_0$. An ideal assumption is that the proton
consists of only valence quarks at extremely low $Q_0^2$.
Thus, a naive nonperturbative input of the proton includes merely three valence quarks
\cite{Parisi,Novikov,Gluck,Chen}, which is the simplest initial parton distributions.
All sea quarks and gluons at high $Q^2$ ($>Q_0^2$) are dynamically produced from QCD evolution.
In fact, there are other types of sea quarks at the starting scale,
such as intrinsic sea \cite{Brodsky,Chang}, connected sea \cite{CS1,CS2,CS3}
and cloud sea \cite{Cloud1,Cloud2,Cloud3}.
Nonetheless, the naive nonperturbative input is generally a good approximation,
because other origins of sea quarks are of small contributions.
The naive nonperturbative input with three valence quarks is very natural in quark model.

In our analysis, valence quark distribution functions at $Q_0^2$ are parameterized
to approximate the analytical solution of nonperturbative QCD.
The simplest function form to approximate valence quark distribution
is the time-honored canonical parametrization
$f(x) = A x^B (1-x)^C$ \cite{CTEQ6}. Hence, the simplest parameterization of
the naive nonperturbative input is written as
\begin{equation}
\begin{aligned}
&u_v(x,Q_0^2)=A_u x^{B_u}(1-x)^{C_u},\\
&d_v(x,Q_0^2)=A_d x^{B_d}(1-x)^{C_d}.
\end{aligned}
\label{Parametrization}
\end{equation}
The parametrization above has poles at $x = 0$ and $x = 1$
to represent the singularities associated with Regge behavior at small $x$
and quark counting rules at large $x$.

In quark model, the proton has two up valence quarks and one down valence quark.
Therefore, we have the valence sum rules for the naive nonperturbative input
\begin{equation}
\int_0^1 u_v(x,Q_0^2)dx=2,
\int_0^1 d_v(x,Q_0^2)dx=1.
\label{ValenceSum}
\end{equation}
Since there are no sea quarks and gluons in the naive nonperturbative input,
valence quarks take the total momentum of the proton.
We have the momentum sum rule for valence quarks at $Q_0^2$,
\begin{equation}
\int_0^1 x[u_v(x,Q_0^2)+d_v(x,Q_0^2)]dx=1.
\label{MomentumSum}
\end{equation}

\section{Standard deviations of quark distribution functions}
\label{SecIII}

The confinement of quarks is a basic feature in non-abelian gauge field theory
\cite{Wilson}. Phenomenologically, Cornell potential is successful for describing
heavy quarkonium, which has linear potential at large distance \cite{Eichten1,Eichten2}.
The linear potential is also realized in LQCD \cite{Kawanai,Evans}.
In MIT bag model \cite{Chodos1,Chodos2,DeGrand}, fields are confined to a finite
region of space. Without doubt, valence quarks inside a proton are confined
in a small space region.

According to Heisenberg uncertainty principle, the momenta of quarks
in the proton are uncertain, which have the probability density distributions.
Heisenberg uncertainty principle is
\begin{equation}
\sigma_X\sigma_P \ge \frac{\hbar}{2}.
\label{Uncertainty}
\end{equation}
To avoid misidentification, the capital $X$ in above formula denotes the ordinary
space coordinate, as lowercase $x$ already denotes the Bjorken scaling variable.
Capital $P$ denotes the momentum in $X$ direction.
$\sigma_X$ is the standard deviation of the spacial position of one parton in $X$
direction, and $\sigma_P$ is the standard deviation of momentum accordingly.
In quantum mechanics, the uncertainty relation is $\sigma_X\sigma_P = 0.568\hbar$
for a particle in a one-dimensional box, and $\sigma_X\sigma_P = \hbar/2$ for
quantum harmonic oscillator at the ground state.
In order to constrain the standard deviations of quark momentum distributions,
$\sigma_X\sigma_P = \hbar/2$ is taken for the three initial valence quarks
in our analysis instead of $\sigma_X\sigma_P \ge \hbar/2$.

$\sigma_X$ is related to the radius of the proton.
An simple estimation is to transform the sphere proton into a cylinder proton,
which gives $\sigma_X = (2\pi R^3/3)/(\pi R^2)=2R/3$,
with $R=\sqrt{<r_p^2>}$ is charge radius of the proton.
Proton charge radius is precisely measured in muonic hydrogen Lamb shift experiments,
which is obtained to be 0.841 fm \cite{Pohl,Antognini}.
$\sigma_X$ of each up valence quark is divided by
$2^{1/3}$ for there are two up valence quarks sharing the same space region.
The confinement space region for up valence quark is half of the total confinement space.
This is an assumption we proposed, not the Pauli blocking principle.
The two up valence quarks have positive electric charges, therefore,
it is very hard for them approaching each other closely.
Consequently, we have $\sigma_{X_d}=2R/3$ and $\sigma_{X_u}=2R/(3\times 2^{1/3})$.

Bjorken variable $x$ is the momentum fraction one parton takes of the proton
momentum in the quark parton model.
Therefore, we define the standard deviation of $x$ at extreme low resolution
scale $Q_0^2$ as
\begin{equation}
\sigma_x = \frac{\sigma_P}{M_p}.
\label{xDeviation}
\end{equation}
$M_p$ is the mass of the proton, which is 0.938 GeV \cite{pdg}.
Natural unit is used in all the calculations of this work.
Finally, constraints for valence quark distributions from QCD confinement
and Heisenberg uncertainty principle are expressed as follows:
\begin{equation}
\begin{aligned}
&\sqrt{<x_u^2>-<x_u>^2}=\sigma_{x_u},\\
&\sqrt{<x_d^2>-<x_d>^2}=\sigma_{x_d},\\
&<x_u>=\int_0^1 x\frac{u_v(x,Q_0^2)}{2}dx,\\
&<x_d>=\int_0^1 xd_v(x,Q_0^2)dx,\\
&<x_u^2>=\int_0^1 x^2\frac{u_v(x,Q_0^2)}{2}dx,\\
&<x_d^2>=\int_0^1 x^2d_v(x,Q_0^2)dx.
\end{aligned}
\label{xDeviation}
\end{equation}

\section{Maximum entropy method}
\label{SecIV}

From above analysis, we do know a lot of information about the valence quark
distributions, but we still cannot get the exact distributions.
By applying maximum entropy principle, we can find the most reasonable valence quark
distributions from the testable information which are the constraints discussed above.
The generalized information entropy of valence quarks is defined as
\begin{equation}
\begin{aligned}
S=&-\int_0^1 [ 2 \frac{u_v(x,Q_0^2)}{2}ln(\frac{u_v(x,Q_0^2)}{2}) \\
  &+ d_v(x,Q_0^2)ln(d_v(x,Q_0^2)) ] dx.
\end{aligned}
\label{EntropyFor}
\end{equation}
The best estimated nonperturbative input will have the largest entropy.
Valence quark distributions are assigned by taking the maximum entropy.

\begin{figure}[htp]
\begin{center}
\includegraphics[width=0.45\textwidth]{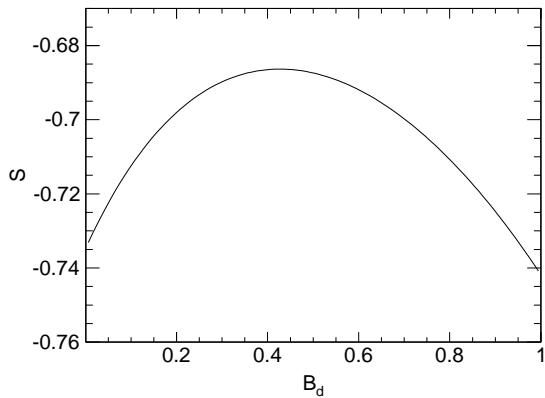}
\caption{
Information entropy $S$ is plotted as a function of the parameter $B_d$.
}
\label{Entropy}
\end{center}
\end{figure}

With constraints given by Equations (\ref{ValenceSum}), (\ref{MomentumSum}) and (\ref{xDeviation}),
there is only one free parameter left for the parameterized naive nonperturbative input.
We take $B_d$ as the only free parameter.
Fig. \ref{Entropy} shows the information entropy of valence quark distributions
of the proton at the starting scale as a function of the parameter $B_d$.
By taking the maximum of the entropy, $B_d$ is optimized to be 0.427.
The corresponding valence quark distributions are
\begin{equation}
\begin{aligned}
&u_v(x,Q_0^2)=4.589x^{0.095}(1-x)^{1.000},\\
&d_v(x,Q_0^2)=7.180x^{0.427}(1-x)^{2.456}.
\end{aligned}
\label{InitialValence}
\end{equation}

\section{Results}
\label{SecV}

By performing Dokshitzer-Gribov-Lipatov-Altarelli-Parisi (DGLAP) evolution
\cite{Dokshitzer,Gribov,Altarelli}, valence quark distributions at high scale
can be determined with the obtained input in Equation (\ref{InitialValence}).
There are only three valence quarks in the proton.
Higher twist corrections to DGLAP equation for valence evolution are small,
for the density of valence quark is not big.
With DGLAP equation, the obtained naive nonperturbative input can be tested
with the experimental measurements at high $Q^2$.
In this work, we use leading order (LO) and next-to-next-to-leading
order (NNLO) evolution. We get the specific starting scale $Q_0^2=0.064$ GeV$^2$
for LO evolution (with $\Lambda_{QCD}=0.204$ GeV for f=3 flavors),
by using QCD evolution for the second moments of the valence quark distributions
\cite{MomentEvo} and the measured moments of the valence quark distributions
at a higher $Q^2$ \cite{GRV98}. This energy scale is very close to the starting
scale for bag model PDFs which is $0.0676$ GeV$^2$ \cite{Steffens}.
The running coupling constant $\alpha_s$
and the quark masses are the fundamental parameters of perturbative QCD.
The running coupling constant for LO evolution we choose is
\begin{equation}
\frac{\alpha_s(Q^2)}{4\pi}=\frac{1}{\beta_0ln(Q^2/\Lambda^2)},
\label{CouplingConst}
\end{equation}
in which $\beta_0=11-2f/3$ and $\Lambda^{3,4,5,6}_{LO}=204, 175, 132, 66.5$ MeV \cite{GRV98}.
For the $\alpha_s$ matchings, we take $m_c=1.4$ GeV, $m_b=4.5$ GeV, $m_t=175$ GeV
for LO evolution. For the NNLO DGLAP evolution, we use the modified Mellin transformation
method by CANDIA \cite{candia}, with $\alpha_s(M_z^2)=0.1155$ and $m_c=1.43$ GeV, $m_b=4.3$ GeV,
$m_t=175$ GeV. The starting scale for NNLO evolution we choose is $Q_0^2=0.22$ GeV$^2$,
which is close to $\Lambda^2_{f=3,NNLO}=0.2$ GeV$^2$. In the NNLO evolution, we have
$\alpha_s(Q_0^2)/(2\pi)=0.3$.

\begin{figure}[htp]
\begin{center}
\includegraphics[width=0.45\textwidth]{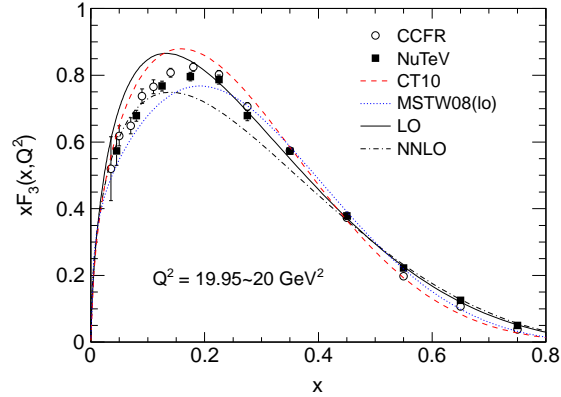}
\caption{
Comparisons of our predicted structure function $xF_3$
(solid and dot-dashed lines) with experimental data from NuTeV
(squares) \cite{NuTeV} and CCFR (open circles) \cite{CCFR}.
Only statistical errors of the experimental data are plotted.
Results of CT10 (dashed line) \cite{CT10} and MSTW08(LO) (dotted line)
\cite{MSTW} from global fit are also shown here.
}
\label{xF3}
\end{center}
\end{figure}

The isoscalar structure function $xF_3$ from neutrino and antineutrino scattering
data provides valuable information of valence quark distributions.
The connection between $xF_3$ and valence quark distributions is given
by $xF_3(x,Q^2)=xu_v(x,Q^2)+xd_v(x,Q^2)$. Our predicted $xF_3$ as a function
of $x$ at high $Q^2$ is shown in Fig. \ref{xF3}, compared with results from
NuTeV and CCFR experiments. The predicted $xF_3$ is in excellent agreement with
the experimental data in large $x$ region ($x>0.3$). On the whole, The LO and NNLO results
are consistent with the experiments except for a small discrepancy around $x=0.1$
and around $x=0.2$, respectively. CT10 and MSTW08(LO) data sets of QCD global analysis
are also plotted in the figure. Our predicted $xF_3$ is close to that from CT10 and MSTW08(LO).

\begin{figure}[htp]
\begin{center}
\includegraphics[width=0.45\textwidth]{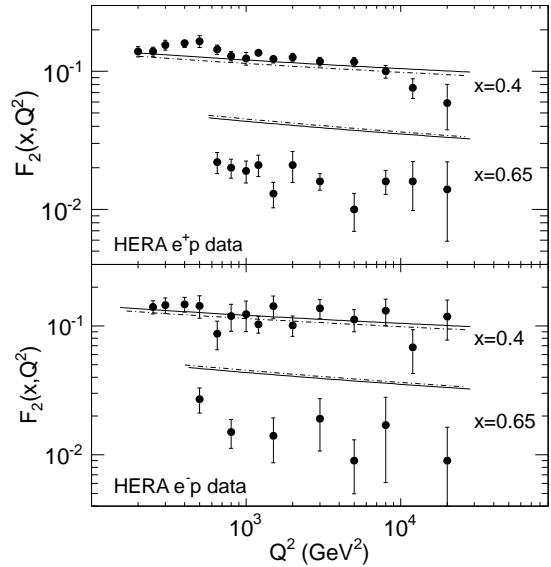}
\caption{
Solid lines and dot-dashed lines are LO and NNLO predictions, respectively.
The combined HERA data  \cite{HERA} are shown in circles.
Errors shown in the plot are the total experimental uncertainties.
Our predicted $F_2$ are from valence contribution only,
assuming sea quarks are negligible at large $x$.
}
\label{F2}
\end{center}
\end{figure}

Structure function $F_2$ plays quite a significant role in determining PDFs,
for it is related to quark distributions directly.
As we know, valence quarks dominate in large $x$ region.
Therefore, $F_2$ at large $x$ is mainly from contributions of valence quarks.
By assuming there are no sea quarks at $x\ge 0.4$, the calculated $F_2$
as a function of $Q^2$ are shown in Fig. \ref{F2}, compared with recent result
from HERA \cite{HERA}. Basically, our predicted $F_2$ are consistent
with the $e^{\pm}p$ neutral-current DIS data.

\begin{figure}[htp]
\begin{center}
\includegraphics[width=0.45\textwidth]{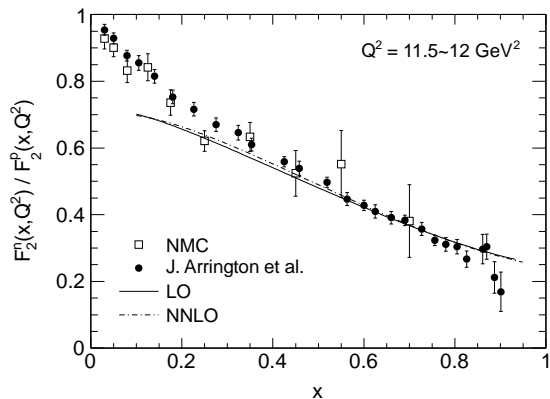}
\caption{
The predicted $F_2$ ratios of neutron to proton (solid and dot-dashed lines)
are shown with experimental data.
Our predicted $F_2$ ratios are calculated without contributions of sea quarks.
NMC data (open squares) is taken from \cite{NMC}.
Detailed analysis data (circles) \cite{Arrington} is from J. Arrington et al.
The plotted errors of experimental data are the total uncertainties.
}
\label{F2nOverF2p}
\end{center}
\end{figure}

Structure function ratio $F_2^n/F_2^p$ is sensitive to both up and down quark distributions.
In large $x$ region, it is mainly related to the up and down valence quark distributions.
Under the assumption of isospin symmetry between the proton and the neutron,
up valence quark distribution in the proton is identical with down valence quark
distribution in the neutron.
Fig. \ref{F2nOverF2p} shows the predicted structure function ratios $F_2^n/F_2^p$
from valence contribution only. Sea quarks are ignored in the calculation.
Experimental results from NMC \cite{NMC} and J. Arrington et al. \cite{Arrington}
are also shown in the figure. Data from J. Arrington et al. are detailed analysis
of previous experimental data within the framework of relativistic quantum mechanics
for the deuteron structure. Our result is in excellent agreement with the
experimental data in the large x region.

\begin{figure}[htp]
\begin{center}
\includegraphics[width=0.45\textwidth]{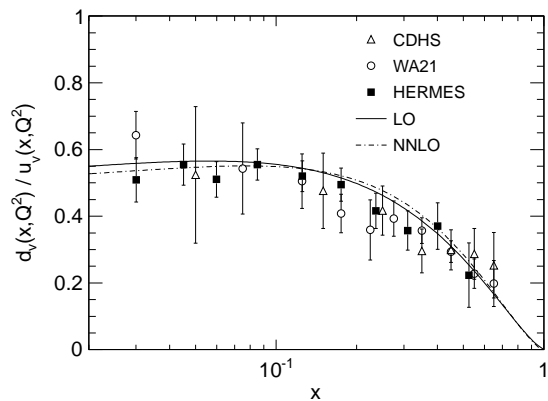}
\caption{
Comparisons of our predicted $d_v/u_v$ ratios (solid and dot-dashed lines)
with experimental results from CDHS (open triangle) \cite{CDHS}, WA21 (open circle)
\cite{WA21} and HERMES (squares) \cite{HERMES}.
The plotted errors are the total errors.
Our predicted ratios are at $Q^2=4$ GeV$^2$.
HERMES data is at mean $Q^2=2.4$ GeV$^2$.
$Q^2$ of CDHS data varies from 3.3 to 42.9 GeV$^2$;
$Q^2$ of WA21 data varies from 3.4 to 36.5 GeV$^2$;
}
\label{DvOverUv}
\end{center}
\end{figure}

Up and down valence quark ratios $d_v/u_v$ are extracted in neutrino DIS and charged
$\pi$ semi-inclusive DIS processes. Our predicted $d_v/u_v$ ratios are shown in
Fig. \ref{DvOverUv} with experimental results from CDHS \cite{CDHS}, WA21 \cite{WA21}
and HERMES \cite{HERMES}. Predicted $d_v/u_v$ ratios at $Q^2=$ 4 GeV$^2$ are
plotted in the figure. $d_v/u_v$ ratios have weak $Q^2$-dependence.
The predicted $d_v/u_v$ ratios agree well with the experimental data.

\begin{figure}[htp]
\begin{center}
\includegraphics[width=0.45\textwidth]{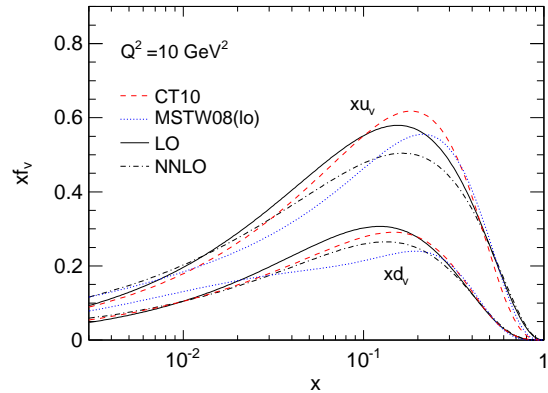}
\caption{
Comparisons of our predicted valence quark momentum distributions
(solid and dot-dashed lines) with global QCD fit CT10
(dashed lines) \cite{CT10} and MSTW08(LO) (dotted lines) \cite{MSTW}.
}
\label{Valence}
\end{center}
\end{figure}

Fig. \ref{Valence} shows the comparisons of our predicted up and down valence quark
momentum distributions, multiplied by $x$, at $Q^2=10$ GeV$^2$ with the global fits
from CT10 \cite{CT10} and MSTW08(LO) \cite{MSTW}.
In general, our obtained up and down valence quark momentum distributions are
consistent with the popular parton distribution functions from QCD global analysis.

\section{Discussions and summary}
\label{SecVI}

Valence quark distributions are given by the maximum entropy method.
This is an interesting attempt of determining the parton distribution functions
using a new method instead of the conventional global fit.
The obtained valence quark distributions are consistent with the experimental
observations from high energy lepton probe and PDFs from global analysis.
The determined valence quark distributions are reasonable, and can be used
for making theoretical predictions. Furthermore, if we make a more complicated
parameterization for the nonperturbative input and include more constraints,
the result possibly becomes better. The Gross-Llewellyn Smith sum rule \cite{GLS1,GLS2,GLS3}
could be the further constraint, which would provide more information on
valence quark distributions at high $Q^2$. The Ellis-Jaffe sum rule
\cite{Ellis-Jaffe,Ellis-Jaffe2} could be practically useful to
constrain the polarized PDFs.

Determining valence quark distributions from maximum entropy method
helps us to understand the primary aspects of the nucleon structure,
and to search for more details of the nucleon. Firstly,
our analysis shows that the origin of PDFs at high $Q^2$ is mainly the three
valence quarks. A simple and naive nonperturbative input is introduced,
and obtained, though it is just an approximation of the complex proton.
Secondly, the basic features of valence quark distributions are related to the
classic quark model assumption, the radius of proton and the mass of proton.
Thirdly, the equation of the uncertainty relation for valence quarks is taken
as the relation for quantum harmonic oscillator at the ground state.
The uncertainty of the momentum could be a little larger.
With the uncertainty being 10\% larger, the obtained prediction becomes a little worse
compared to experiments. More detailed study of the confinement potential will
put on more accurate constraints to the uncertainty relation.
Finally, the time-honored canonical parametrization scheme for valence quarks
is very simple, but acceptable.

Maximum entropy method is applicable for obtaining details of interest
with least bias in situations where detailed information is not given.
It is difficult to calculate the radius and mass of the proton from nonperturbative QCD.
LQCD cannot acquire the detailed information of nucleon structure so far.
However, we do know the radius and mass of the proton from measurements
in experiments and the confinement of quarks in QCD theory.
With these experimental observations and some assumptions, the best estimate
of valence quark distributions are obtained from maximum entropy method.
Because of the simplicity, this method can be easily applied to other types of PDFs,
such as polarized PDFs, generalized parton distributions and Transverse
momentum dependent PDFs. Maximum entropy method is particularly useful for digging
reasonable results in situations where relatively little information
from QCD calculation is given.

\noindent{\bf Acknowledgments}:
This work was supported by the National Basic Research Program of China (973 Program)
2014CB845406, the National Natural Science Foundation of China under
Grant Number 11175220 and Century Program of Chinese Academy of Sciences Y101020BR0.

\end{document}